\documentclass[twocolumn,prb,color,psfig,superscriptaddress]{revtex4}
\usepackage{graphicx}
\usepackage{epsfig}
\def\Li{LiCu$_2$O$_2$}
\def\LiV{LiCuVO$_4$}
\def\beq{\begin{equation}}
\def\eeq{\end{equation}}

\begin{document}
\title{Nonrelativistic Multiferroicity    in  the Nonstoichiometric Spin s=1/2 Spiral Chain Cuprate  \Li}
\author{A.S. Moskvin}
\affiliation{Ural State University, 620083 Ekaterinburg,  Russia}
\author{Yu.D. Panov}
\affiliation{Ural State University, 620083 Ekaterinburg,  Russia}
\author{S.-L. Drechsler}
\affiliation{Leibniz Institut f\"ur Festk\"orper- und Werkstofforschung
Dresden, D-01171, Dresden, Germany}
\date{\today}


\begin{abstract}

We argue for a  recently observed puzzling multiferroic behavior in s=1/2 1D  chain cuprate  \Li \, with edge-shared arrangement of CuO$_4$ plaquettes and incommensurate spiral spin ordering can be consistently explained if one takes into account the nonrelativistic exchange-induced electric polarization on the Cu$^{2+}$ centers substituting for the positions native for  the Cu$^{1+}$-ions. These substituent centers are proved to be an effective probe of the spin incommensurability and magnetic field effects. 
\end{abstract}

\maketitle

Recent observations of multiferroic behaviour concomitant the incommensurate  spin spiral ordering in  chain cuprates LiCuVO$_4$ \cite{Naito,Naito1,Schrettle} and LiCu$_2$O$_2$   \cite{Cheong} challenge the multiferroic community. At first sight, these  cuprates seem to  be  prototypical examples of 1D spiral-magnetic ferroelectrics revealing the $relativistic$ mechanism of  "ferroelectricity caused by spin-currents"\cite{Katsura1} with the textbook expression for the $uniform$ polarization induced by a spin spiral    with the wave vector $\bf Q$: 
$
{\bf P}\propto\left[{\bf e}_3\times {\bf Q}\right],
$ 
where ${\bf e}_3$ is a vector orthogonal to the spin spiral plane\cite{Mostovoy} or 
$	{\bf P}_{ij}\propto\left[ {\bf R}_{ij}\times \left[{\bf S}_{i}\times {\bf S}_{j}\right]\right]$, where ${\bf R}_{ij}$ denotes the vector connecting the two sites  and $\left[{\bf S}_{i}\times {\bf S}_{j}\right]$ is a local spin current.\cite{Katsura1}
However, the both systems reveal a mysterious behavior with conflicting results obtained by different groups. Indeed, Yasui {\it et al.}\cite{Naito1}  claim the LiCuVO$_4$ reveals  clear deviations from the predictions of spin-current models\cite{Mostovoy,Katsura1} while Schrettle {\it et al.}\cite{Schrettle}  assure of its applicability. In contrast to LiCuVO$_4$, the LiCu$_2$O$_2$ shows up a behavior which is obviously counterintuitive within the framework of spiral-magnetic ferroelectricity.\cite{Cheong} It is worth noting that at variance with Park {\it et al.}\cite{Cheong}, Naito {\it et al.}\cite{Naito} have not found any evidence for  ferroelectric anomalies in LiCu$_2$O$_2$. Such a discrepancy one observes in microscopic model approaches as well. The relativistic LSDA calculations\cite{Xiang} seemingly explain the LiCuVO$_4$ data\cite{Schrettle} but fail in case of LiCu$_2$O$_2$. However, a detailed analysis of relativistic effects for  the system of $e_g$-holes in a  perfect chain structure 
of edge-shared CuO$_4$ plaquettes as in LiCuVO$_4$ and LiCu$_2$O$_2$ shows that the in-chain spin current does not produce an electric polarization because of an exact cancellation of two Cu-O-Cu paths.\cite{mechanism} Moreover, recently we have shown\cite{EPL}  that the multiferroicity in LiCuVO$_4$ may have nothing to do with $relativistic$ effects and can be consistently explained, if  the $nonrelativistic$ exchange-induced electric polarization on the out-of-chain Cu$^{2+}$ centers substituting for Li-ions in LiVCuO$_4$ is taken into account. Below we argue that  a similar mechanism which  takes into account the  exchange-induced electric polarization on the Cu$^{2+}$ centers, substituting unexpectedly for  Cu$^{1+}$-ions, is at work in LiCu$_2$O$_2$.
 
 LiCu$_2$O$_2$ is orthorhombic   mixed-valent compound with copper ions  in the Cu$^{2+}$ and   Cu$^{1+}$ valence states.\cite{Hibble}
 The unit cell contains four magnetic Cu ions belonging to two pairs of CuO$_2$ chains formed by edge-shared Cu$^{2+}$O$_4$ plaquettes running along the 
crystallographic $b$-axis and linked  by  the LiO$_5$ double chains. Alternating double parallel chains, containing either Li or Cu atoms,
form the sheets which are interconnected by Cu$^{1+}$ in O-Cu-O dumbbells.
  
 The first experimental evidence of magnetic incommensurability in \Li\, was obtained independently by Gippius {\it et al.}\cite{Gippius}  and Masuda {\it et al.}\cite{Masuda}  from $^{6,7}$Li NMR  and neutron diffraction measurements, respectively. Any spins related by a translation along the $\bf c$ axis and $\bf a$ axis are parallel and
antiparallel to each other, respectively.  A good fit to neutron diffraction data was obtained with all spins confined to the  $ab$  crystallographic plane\cite{Masuda} thus forming $ab$-plane spin spirals running along $\bf b$ axis:${\bf S}(y)= S(\cos\theta ,\sin\theta,0)$, where  $\theta=qy+\alpha$,   $\alpha$ is a phase shift. 
\begin{figure}[h]
\includegraphics[width=8.5cm,angle=0]{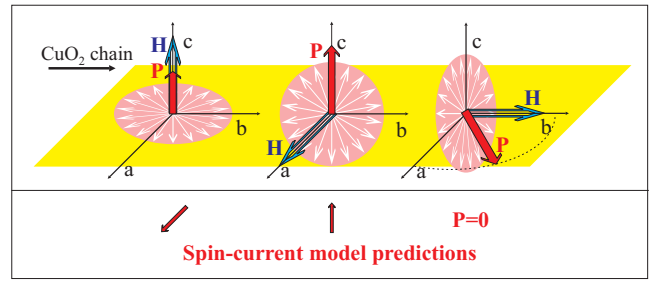}
\caption{Direction of ferroelectric polarization in \Li \, for different spin spiral plane orientation. }  \label{fig1}
\end{figure}   
 Park {\it et al.}\cite{Cheong} have found that the incommensurate  spin ordering in  LiCu$_2$O$_2$ below T$_N\approx$ 23 K is accompanied by a ferroelectric transition with a puzzling anisotropy and field dependence which are reproduced schematically in Fig. 1. First of all, the electric polarization in zero field is directed along the $\bf c$ axis implying in accordance with the concept of  spin current induced  ferroelectricity  that the spiral spins lie  in the $bc$-plane in sharp contrast with earlier neutron diffraction data.\cite{Masuda} When a magnetic field applied along the $\bf b$ axis (see Ref.\onlinecite{Cheong} for the making use  of $a,b$ notations in ab-twinned crystal), $P_c$ decreases and $P_a$ increases, implying that the Cu$^{2+}$ spin spiral plane flips from the $bc$ to $ab$ plane,  resulting in a flip of the polarization  from the $c$ to $a$ axis. It is expected that ${\bf h}\parallel {\bf c}$ may flip the spiral plane from the $bc$ to $ab$ plane, so that $\bf P$ may flip from the $\bf c$ to $\bf a$ axis with $h_c$.  However, this is completely in contrast with the observations\cite{Cheong} that $h_c$ enhances
$P_c$ and $h_b$ is the one inducing the $\bf P$ flip from the $\bf c$ to $\bf a$ axis. The appearance of $P_a$ with $h_a$ is also counterintuitive within the framework of the relativistic spiral-magnetic ferroelectricity.\cite{Mostovoy,Katsura1} These unexpected magnetic field effects raise doubts about the validity of the scenario of  relativistic spin curent   spiral-magnetic ferroelectricity and point to another, probably the out-of-spin-chain  origin of the magnetoelectric coupling.
\begin{figure}[t]
\includegraphics[width=8.5cm,angle=0]{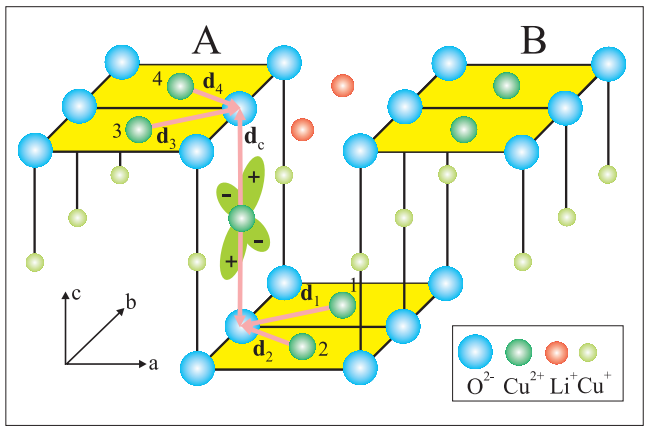}
\caption{An  idealized view of crystal structure of \Li (upper bilayer). A "left" site impurity center with Cu$^{2+}$-ion substituted for Cu$^{1+}$-ion is inbetween upper and lower CuO$_2$ chains from the same unit cell. Shown is the hole density distribution in $d_{yz}$ orbital. The exchange induced dipole moments are shown by arrows.} \label{fig2}
\end{figure}
In this connection it is worth noting that the thermogravimetric analysis revealed  the \Li \,samples had a lower content of Cu
ions than follows from the stoichiometric formula.\cite{Masuda} Chemical disorder and a Cu deficiency by asmuch as x =  16\% are  inherently present. The "surplus" Li$^+$ ions in \Li \, occupy Cu$^{2+}$ sites, due to a good match of ionic radii (0.68 and 0.69 \AA, respectively).
The charge compensation requires that the introduction of nonmagnetic Li$^+$ ions into the double chains is accompanied by a transfer of the S=1/2-carrying Cu$^{2+}$ ions onto the Cu$^{+}$ interchain sites.\cite{Masuda} At first sight it seems improbable because of different coordination preferences. However, the actual coordination of the native Cu$^{+}$ interchain site approaches most likely to an axially distorted square, or rhombic  coordination due to an extremely small inter-dumbbell separation ($d\approx 2.86 \AA$) as compared with other  O-Cu$^{+}$-O dumbbell bearing compounds (e.g. YBa$_2$CuO$_6$, $d\approx 3.8 \AA$).\cite{Pisarev-LiCuO} In other words, the Zhang-Rice singlet within the CuO$_2$ chains becomes unstable with respect to a hole transfer to one of neighboring Cu$^+$ sites. Details of this instability will be discussed elsewhere. 
If the doped hole would be remain in the CuO$_2$ chains,  dimer-type effects as in other hole doped chains would be observed experimentally. Also the spiral state observed in the neutron diffraction would be strongly disturbed by the presence of these holes.
What is the ground state of the single hole configuration of Cu$^{2+}$ ion in the native Cu$^{+}$ interchain sites? Purely electrostatic arguments made within the framework of the point charge model, supported by account for Cu 3d-O 2p covalency,  point to a competition of $d_{z^2}$ and $d_{yz}$ orbitals while strong intra-atomic $s$-$d_{z^2}$ hybridization singles out the 
$d_{yz}$ orbital to be a main candidate for the ground state. 
The Cu$^{2+}$ substituents in  native  Cu$^{1+}$ positions form  strongly polarizable entities which electric polarization due to a parity-breaking exchange interaction\cite{TMS} with Cu$^{2+}$ spin spirals explains all the puzzles observed by Park {\it et al.}\cite{Cheong} This unconventional exchange coupling can be easily illustrated for, e.g.,  the one-particle (electron/hole) center in a crystallographically centrosymmetric position of a magnetic crystal when all the particle states can be of definite spatial parity, even (g) or odd (u), respectively. Having in mind the 3d centers we'll assume an even-parity ground state $|g\rangle$. For simplicity we restrict ourselves by only one excited odd-parity state  $|u\rangle$. The exchange coupling with the surrounding spins can be written as follows:
\begin{equation}
	{\hat V}_{ex}=\sum_n{\hat I}({\bf R}_n)({\bf s}\cdot {\bf S}_n),
\end{equation}
where ${\hat I}({\bf R}_n)$ is an orbital operator with a matrix
\begin{equation}
{\hat I}({\bf R}_n)=\pmatrix{I_{gg}({\bf R}_n)&I_{gu}({\bf R}_n)\cr
I_{ug}({\bf R}_n)&I_{uu}({\bf R}_n)\cr}.
\end{equation}
The parity-breaking off-diagonal part of the exchange coupling can lift the center of symmetry and mix $|g\rangle$ and $|u\rangle$ states 
giving rise to a nonzero electric dipole polarization of the ground state
\begin{equation}
{\bf P}=2c_{gu}\langle g|e{\bf r}|u\rangle =\sum_{n}{\bf \Pi}_n({\bf s}\cdot {\bf S}_n)\, 
\end{equation}
with
$
{\bf \Pi}_n =	2I_{gu}({\bf R}_n)\langle g|e{\bf r}|u\rangle /\Delta_{ug}$
($\Delta _{ug}=\epsilon_u-\epsilon_g$). 

Strictly speaking, the parity-breaking  exchange coupling of native Cu$^{2+}$ center in CuO$_2$ chain (hole ground state $|g\rangle \propto d_{xy}$) with neighbouring Cu$^{2+}$ substituent (hole ground state $|g\rangle \propto d_{yz}$) will result in the $ab$-plane electric polarization of CuO$_4$ chain plaquettes and the $c$-axis polarization of the Cu$^{2+}$ substituent. 

Unit cell of \Li \, contains two types (left and right) of native Cu$^{1+}$-positions (see Fig.\,\ref{fig1}) with four neighbouring Cu$^{2+}$ centers in the two CuO$_2$ chains. Within the framework of our model the both  "left" A-type and  "right" B-type substituent centers differ by the spin spiral phase shift  $\alpha =\pi /2$ and $\alpha =-\pi /2$ with respect to the lower chain, and by orientation of the generated electric dipole moments: $d_a(A)=-d_a(B)=d/\sqrt{2}$, $d_b(A)=d_b(B)=d/\sqrt{2}$, $d_c(A)=d_c(B)=d_c$.
Here we ignore the weak influence of the adjacent chain in the third CuO$_2$ chain belonging to the adjacent bilayer. According to LDA calculation there is practically no hybridization with that chain.\cite{Gippius}

To describe different configurations of the spin neighbourhood for a Cu$^{2+}$ substituent (see Fig.\,\ref{fig2}) we introduce four basic vectors similarly to conventional ferro- and antiferromagnetic vectors as follows:
$$
{\bf F}(y)=[{\bf S}_1	+{\bf S}_2+{\bf S}_3+{\bf S}_4]; \, 
{\bf G}(y)=[{\bf S}_1	-{\bf S}_2+{\bf S}_3-{\bf S}_4];
$$
$$
{\bf A}(y)=[{\bf S}_1	+{\bf S}_2-{\bf S}_3-{\bf S}_4]; \,
{\bf C}(y)=[{\bf S}_1	-{\bf S}_2-{\bf S}_3+{\bf S}_4] \, 
$$
with a kinematic constraint: $({\bf F}\cdot{\bf A})$=$({\bf C}\cdot{\bf G})$=0 valid for two identical spirals irrespective of their phase shift. Then the electric polarization induced by the parity-breaking exchange coupling of  Cu$^{2+}$ substituent with a complete set of four neighboring in-chain Cu$^{2+}$ ions 1-4 (see Fig.\,\ref{fig1}) can be written as follows:
$$
P_{a,c}=d_{a,c}({\bf s}\cdot {\bf A});\, P_b=d_{b}({\bf s}\cdot {\bf C})\, .
$$
Spin polarization of Cu$^{2+}$ substituent spin can be easily found within the framework of a weak coupling approximation, if one take the most general form of the impurity-spiral ground state ($gg$) exchange interaction
\beq
V_{sS}=\sum_{i=1-4}\hat{\bf s}{\bf \stackrel{\leftrightarrow}{I}(i)} \,\hat{\bf S}_i =(\hat{\bf s}\cdot \hat{\bf H}_0)\, ,\label{1}
\eeq
where $\hat{\bf H}_0$ is an effective magnetic field, acting on the Cu$^{2+}$ substituent, $I_{\alpha\alpha}(i)=I_{\alpha\alpha},I_{xz}(i)=I_{xz};
I_{xy}(1)=-I_{xy}(2)=I_{xy}(3)=-I_{xy}(4)=I_{xy};
I_{zy}(1)=-I_{zy}(2)=I_{zy}(3)=-I_{zy}(4)=I_{zy}.
$
are a symmetric matrix of the exchange integrals. Thus for the effective field we obtain
\beq
{\bf H}_0(y)={\bf \stackrel{\leftrightarrow}{I}_F}{\bf F}+{\bf \stackrel{\leftrightarrow}{I}_G}{\bf G}
\eeq
with
$$
{\bf \stackrel{\leftrightarrow}{I}_F} =\pmatrix{I_{xx} & 0 & I_{xz} \cr 0 & I_{yy} & 0 \cr I_{xz} & 0 & I_{zz}\cr}; \,\,
{\bf \stackrel{\leftrightarrow}{I}_G}=\pmatrix{0 & I_{xy} & 0 \cr I_{xy} & 0 & I_{zy} \cr 0 & I_{zy} & 0\cr}\, .
$$

We start with the $ab$-plane spiral ordering of Cu$^{2+}$ spins in the CuO$_2$ chains of \Li \,, which  deduced from neutron diffraction data in zero external field\cite{Masuda} and assume $T=0$.
 For zero external magnetic field or for a field directed along the $\bf c$ axis and for $\alpha =\pm\pi /2$ the electric polarization of the $y$-th Cu$^{2+}$ substituent center  oscillates as follows
\beq
P_c(y)= \frac{8d_cuS^2}{H(y)}[(I_{xx}-I_{yy})u\cos(2qy)\pm 2I_{xy}v\sin(2qy)]\, , \label{P_c(y)}
\eeq
where $u=\cos(\frac{qb}{2}), v=\sin(\frac{qb}{2})$, and for $h=0$
$$
H(y)= 2\sqrt{2}S[(I^2_{xx}+I^2_{yy}+I^2_{zx})u^2+(2I^2_{xy}+I^2_{zy})v^2)
$$
$$
+2((I_{xx}+I_{yy})I_{xy}+I_{zx}I_{zy})uv\cos(2qy)
$$
\beq
\mp (((I_{xx}+I_{yy})(I_{xx}-I_{yy})+I^2_{zx})u^2-I^2_{zy}v^2)\sin(2qy)]^{\frac{1}{2}}\, . \label{Hab-plane}
\eeq 
First, it should be noted that the both  "left" A-type ($\alpha =\pi /2$) and  "right" B-type ($\alpha =-\pi /2$) substituent positions contribute equally to a macroscopic polarization $P_c$. On the other hand, it means that $P_a$ vanishes due to an exact compensation of A-type and B-type contributions since $d_a(A)=-d_a(B)$. For $P_b$ we arrive at  a strict cancellation of the   net electric polarization given $\alpha =\pm\pi /2$ due to opposite signs of the antisymmetric part of the effective field. Moreover, this cancellation hold itself also under an external magnetic field irrespective of its direction.
Second, we  note that  a nonzero electric polarization for the substituent center 12-Cu$^{2+}_A$-34 can be related only with the anisotropic substituent-spiral exchange coupling.
The net polarization $\left\langle P_c(y)\right\rangle$ seems to be rather weak because of several reduction effects: i) the existence of non-compensated non-oscillatory contribution of isotropic exchange to the effective magnetic field (\ref{Hab-plane}); ii) a quadratic or cubic dependence of $\left\langle P_c(y)\right\rangle$ on the exchange anisotropy parameters. 
\begin{figure}[b]
\includegraphics[width=8.5cm,angle=0]{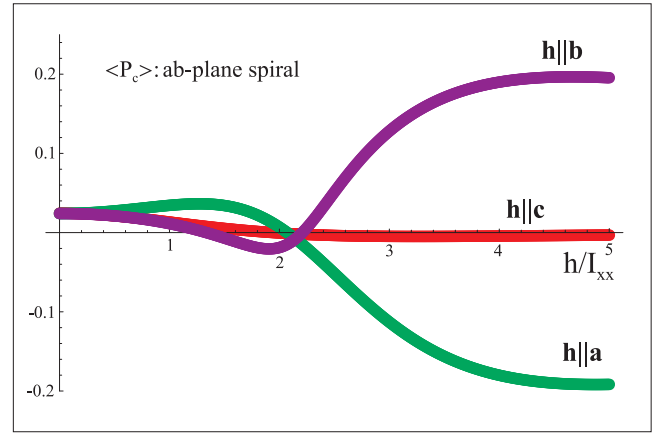}
\caption{The  field dependence of  $\left\langle P_c(y)\right\rangle$ (in units of $d_c$) for $ab$-plane spiral:  $I_{xy}/I_{xx}=0.4$ }  \label{fig3}
\end{figure}
In order to demonstrate the role of the anisotropic exchange we adopt a relation between the anisotropy parameters predicted by a simple nearest-neighbor magneto-dipole model: $(I_{xx}-I_{yy})= 0$, $(I_{zx}= I_{zy})= \sqrt{2}I_{xy}$. The dependence of $\left\langle P_c(y)\right\rangle$ on the ratio $\delta =I_{xy}/I_{xx}$ appears to be  strongly nonlinear, being approximately $\propto\delta ^3$ for a small anisotropy. Only a strong anisotropy $\delta \sim 1$ provides the magnitudes of $\left\langle P_c(y)\right\rangle$ comparable with that of  $\left\langle P_a(y)\right\rangle$ in \LiV.
The typical field dependence of $\left\langle P_c(y)\right\rangle$   is shown in Fig.\,\ref{fig3} given $qb/2=0.172\pi$ which corresponds to a pitch angle $\approx 62^{\circ}$.\cite{Masuda,Drechsler1}

A magnetic field ${\bf h}\parallel {\bf a}$  induces in \Li \, a $ab$-$bc$ spin-flop transition to the phase with a $bc$-plane spiral ordering.
Interestingly, that irrespective of the field direction a $bc$-plane spin spiral 
ordering, similarly to that of $ab$-plane one, supports only a $c$-axis orientation of both local and net electric polarizations, which  expressions can be easily obtained from  
their $ab$-axis counterparts, if one makes the  interchange:$h_x\rightarrow h_z$, $I_{xy}\leftrightarrow I_{zy}$.
It is worth noting that at variance with the $ab$-plane spin spiral ordering the $c$-axis orientation of net electric polarization for $bc$-plane spin arrangement agrees with the predictions of the spin current scenario. Thus both the local  ${\bf P}(y)$ and the averaged electric polarizations $\left\langle {\bf P}(y)\right\rangle$ for $ab$ and $bc$ plane spin spirals lie along the $c$-axis even in zero magnetic field. It's quite another matter for the $ac$-plane spin spiral arrangement which can be a result of a spin-flop transition in an external magnetic field directed along $b$-axis. 
Contributions of the A- and B-type centers to $P_a$ and $P_c$ are strictly opposite in  sign, that means their cancellation for $P_c$ and doubling for $P_a$. From the other hand, for the first time, the $P_b$ component of electric polarization appears to be nonzero. Thus, in contrast with two preceding instances both the local  ${\bf P}(y)$ and the averaged electric polarizations $\left\langle {\bf P}(y)\right\rangle$ for $ac$-plane spin spirals lie in $ab$ plane even in zero magnetic field. Moreover, 
we arrive at a simple relation between the $a$- and the $b$-components of the electric polarization:
$
P_b/P_a=-v/u=- \tan(qb/2),
$
\begin{figure}[t]
\includegraphics[width=8.0cm,angle=0]{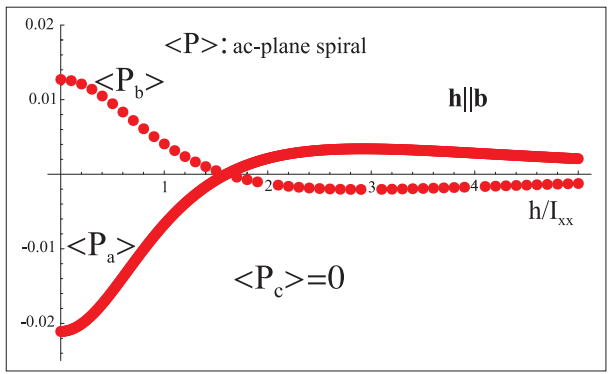}
\caption{The field dependence of  $\left\langle P_{a,b}(y)\right\rangle$ (in units of $d$) for ${\bf h}\parallel {\bf b}$ ($I_{xy}/I_{xx}=0.4$)  for the $ac$-plane spiral.}  \label{fig4}
\end{figure}
that corresponds to $P_b/P_a\approx -0.6$ given the pitch angle $qb\approx 62^{\circ}$.
Fig.\,\ref{fig4} shows the field dependence of  $\left\langle P_{a,b}(y)\right\rangle$ for ${\bf h}\parallel {\bf b}$ ($I_{xy}/I_{xx}=0.4$)  for the $ac$-plane spiral.

The mechanism of impure ferroelectricity we discuss does consistently explain all the puzzles of the magnetoelectric effect observed in \Li \, by Park {\it et al.}\cite{Cheong}(see Fig. 1).
First of all the model explains the $c$-axis direction of the spontaneous electric polarization emerging below the spiral-magnetic ordering temperature within the framework of a dominant $ab$ plane Cu$^{2+}$ spin arrangement, proposed earlier from neutron diffraction data.\cite{Masuda}
We argue that an external field ${\bf h}\parallel {\bf b}$ induces  spin-flop transition with the Cu$^{2+}$ spin spiral plane flipping from the $ab$ to the $ac$ plane accompanied by 
the flipping of net electric polarization ${\bf P}$ from the $c$-axis to the $ab$-plane where the relation imbetween $b$- and $a$-components is determined by the actual pitch angle.
The twin structure observed in $ab$ plane of the \Li \, crystal and ferroelectric domain effects\cite{Cheong} make the field dependence of electric polarization quite complex. 
Indeed, an external field ${\bf h}\parallel {\bf b}$ induces different spin-flop transitions in different twins: $ab$$\rightarrow$$ac$ and $ab$$\rightarrow$$bc$, respectively. Only in the former twins we deal with $P_c$$\rightarrow$$P_{ab}$ flipping of ferroelectric moment, while in the latter twins the polarization remains oriented along the $c$-axis, though having a varied magnitude as compared with the $ab$ plane spin spiral. Such a behavior is observed in experiments by  Park {\it et al.},\cite{Cheong} with a relation between the in-plane components of polarization which is close to a theoretically predicted value 0.6.
At variance with \LiV \, the spontaneous ($h=0$) electric polarization in the low-temperature spiral phase of \Li \, does  depend not only on the pitch angle ($qb$) and
 the relation inbetween the values of exchange anisotropy parameters, but also on the relative magnitude of exchange anisotropy as compared with isotropic exchange. Namely this feature is believed to determine the relatively small magnitude of the multiferroic effect in \Li \, as compared with \LiV.\cite{Naito} 

Thus we conclude that at variance with the $relativistic$ spin current model the $nonrelativistic$ parity breaking exchange induced polarization for the centers formed by Cu$^{2+}$ substituted for Cu$^{1+}$ in nonstoichiometric \Li \,with a simple zero-field  $ab$-plane spiral ordering can be a natural electronic source of multiferroicity found by Park {\it et al.}\cite{Cheong} in this cuprate.

We thank  A. Loidl, U.\ R\"o{\ss}ler, K. D\"orr, R. Kuzian and H. Rosner for discussions.
The  DFG and RFBR Grants (Nos.  06-02-17242, 06-03-90893,  and  07-02-96047)
 are acknowledged for financial support.

\end{document}